\documentclass[]{aastex}

\usepackage{emulateapj5}
\usepackage{onecolfloat}
\usepackage{graphicx} 
\usepackage{fancyheadings} 
\usepackage{ulem}
\usepackage{rotating}
\usepackage{lscape}

\newcommand{\ndla}{71}
\newcommand{\kms}{km~s$^{-1}$ }
\newcommand{\cm}[1]{\, {\rm cm^{#1}}}
\newcommand{\mkms}{{\rm \; km\;s^{-1}}}

\newcommand{\ohi}{$\Omega_g$}
\newcommand{\lya}{Ly$\alpha$}
\newcommand{\nv}{N\,V}
\newcommand{\ovi}{O\,VI}
\newcommand{\N}[1]{{N({\rm #1})}}
\newcommand{\sci}[1]{{\rm \; \times \; 10^{#1}}}
\newcommand{\tskip}{\tablevspace{1pt}}

\begin{document}

\twocolumn[%
\submitted{Accepted to PASP: April 20, 2004}
\title{The SDSS Damped \lya\ Survey: Data Release 1}

\author{Jason X. Prochaska \& Stephane Herbert-Fort}
\affil{UCO/Lick Observatory;
University of California, 1156 High Street, Santa Cruz, CA 95064}
\email{xavier@ucolick.org, shf@ucolick.org}

\begin{abstract}
We present the results from an automated search for damped \lya\
(DLA) systems in the quasar
spectra of Data Release 1 from the Sloan Digital Sky Survey (SDSS-DR1).
At $z\approx 2.5$, this homogeneous dataset has greater statistical 
significance than the previous two decades of research.
We derive a statistical sample of \ndla\ damped \lya\ systems
($>50$ previously unpublished) at $z>2.1$ and measure
H\,I column densities directly from the SDSS spectra.
The number of DLA systems per unit redshift 
is consistent with previous measurements
and we expect our survey has $>95\%$ completeness.
We examine the cosmological baryonic mass density of neutral gas
$\Omega_g$ inferred from the damped \lya\ systems from the SDSS-DR1
survey and a combined sample drawn from the literature.
Contrary to previous results, the $\Omega_g$ values 
do not require a significant correction from 
Lyman limit systems at any redshift.  
We also find that the $\Omega_g$ values for 
the SDSS-DR1 sample do not decline at high redshift
and the combined sample 
shows a (statistically insignificant) decrease only at $z>4$.
Future data releases from SDSS will provide the definitive survey
of DLA systems at $z\approx 2.5$ and will significantly reduce
the uncertainty in $\Omega_g$ at higher redshift.

\keywords{Galaxies: Evolution, Galaxies: Intergalactic Medium, 
Galaxies: Quasars: Absorption Lines}

\end{abstract}
]

\pagestyle{fancyplain}
\lhead[\fancyplain{}{\thepage}]{\fancyplain{}{PROCHASKA AND HERBERT-FORT}}
\rhead[\fancyplain{}{The SDSS Damped \lya\ Survey I}]{\fancyplain{}{\thepage}}
\setlength{\headrulewidth=0pt}
\cfoot{}

\section{Introduction}

It has now been two decades since the inception of surveys
for high redshift galaxies through the signature of damped
\lya\ (DLA) absorption in the spectra of background quasars \citep{wolfe86}.
Owing to large neutral hydrogen column densities $\N{HI}$,
these absorption lines 
exhibit large rest equivalent widths ($W_\lambda > 10$\AA) and show the
Lorentzian wings characteristic of quantum mechanic line-damping.
Through dedicated surveys of high and low redshift quasars with optical
and ultraviolet telescopes,
over 300 damped \lya\ systems have been identified.  These
galaxies span redshifts $z=0$ (the Milky Way, LMC, SMC) to $z=5.5$ 
where the opacity of the \lya\ forest precludes detection
\citep{songaila02}.

Statistics of the DLA systems impact a wide range of topics
in modern cosmology, galaxy formation, and physics.
These include studies on the chemical enrichment of the universe in neutral
gas \citep{pettini94,pro03b}, nucleosynthetic processes
\citep{lu96,phw03}, galactic velocity fields \citep{pw97}, 
the molecular and dust content of young galaxies \citep{vladilo98,ledoux03},
star formation rates \citep{wpg03}, 
and even constraints on temporal evolution of the fine-structure
constant \citep{webb01}. 
Perhaps
the most fundamental measurement from DLA surveys, however, is the evolution
of the cosmological baryonic mass density in neutral gas \ohi\ 
\citep[][ hereafter PMSI03]{storrie00,rao00,peroux03}.
Because the DLA systems dominate the mass density of neutral gas
from $z=0$ to at least $z=3.5$, a census of these absorption systems
determines directly $\Omega_g$.
These measurements express global evolution in the gas which
feeds star formation \citep{pei95,mathlin01} and are an important
constraint for models of hierarchical galaxy formation 
\citep[e.g.][]{spf01,nagamine04a}.

The most recent compilation of damped \lya\ systems surveyed in
a `blind', statistical manner combines the effects
of observing programs using over 10 telescopes, 10 unique instruments,
and the data reduction and analysis of $\approx 10$ different observers
(PMSI03).
In short, the results are derived from a heterogeneous sample of
quasar spectra derived from heterogeneous quasar surveys.
While considerable care has been paid to collate these studies into
an unbiased analysis, it is difficult to assess the completeness
and potential selection biases  of the current sample.
These issues are particularly important when one aims to
address the impact of effects like dust obscuration
\citep{ostriker84,fall93,ellison01}.

In this paper we present the first results in a large survey
for damped \lya\ systems drawn from a homogeneous dataset 
of high $z$ quasars with well-defined selection criteria.
Specifically, we survey the quasar spectra from 
Data Release 1 of
the Sloan Digital Sky Survey  (SDSS-DR1)
restricting our search
to SDSS-DR1 quasars with Petrosian magnitude $r' < 19.5$\,mag.  
The DR1 sample alone (the first of five data releases
from SDSS) offers a survey comparable to 
-- although not strictly independent from -- the efforts of
20 years of work.  We introduce algorithms to automatically identify
DLA candidates in the fluxed (i.e.\ non-normalized) quasar
spectra and perform Voigt profile analyses to confirm and analyze
the DLA sample.  This survey was motivated by a search for
`metal-strong' DLA systems like the $z$=2.626 damped 
\lya\ system toward FJ$0812+32$ \citep{phw03}.  A discussion of the
`metal-strong' survey
will be presented in a future paper (Herbert-Fort et al.\ 2004,
in preparation).  

This paper is organized as follows.  In $\S$~2, we present the
quasar sample and discuss the automatic DLA candidate detection.
In $\S$~3, we present the Voigt profile fits to the full sample.
We present a statistical analysis in $\S$~4 and a summary and concluding
remarks are given in $\S$~5.

\clearpage

\section{Quasar Sample and DLA Candidates}

The quasar sample was drawn from Data Release 1 of the Sloan
Digital Sky Survey to a limiting Petrosian
magnitude of $r' = 19.5$\,mag.
This criterion was chosen primarily to facilitate follow-up 
observations with 10m-class telescopes and it
includes $>60\%$ of all SDSS-DR1 quasars at $z>2$.  With rare exception, 
the fiber-fed SDSS spectrograph provides FWHM~$\approx 150 \mkms$
spectra of each quasar for the wavelength range $\lambda \approx 3800 - 9200$\AA. 
All of the spectra were reduced using the SDSS 
spectrophotometric pipeline
\citep{burles04} and were retrieved from the SDSS data 
archive\footnote{http://www.sdss.org} \citep{abazajian03}.

The first step of a damped \lya\ survey is to establish the 
redshift pathlength available to the discovery of DLA systems.
The minimum starting wavelength
of 3800\AA\ corresponds to $z=2.12$ for the \lya\ transition and this
sets the lowest redshift accessible to this survey.
For each quasar, however, we define a unique starting redshift
$z_{start}$ by identifying 
the first pixel where the median SNR over 20 pixels 
exceeds 4.  
This criterion was chosen to (1) 
minimize the likelihood of identifying noise features as DLA systems;
(2) achieve a high completeness limit;
(3) account for the presence of Lyman limit absorption.
Consistent with previous studies, the ending redshift $z_{end}$
corresponds
to $3000 \mkms$ blueward of \lya\ emission. 
This criterion limits the probability of identifying DLA systems
associated with the quasar which may bias the analysis.

Special consideration is given to quasar spectra which show
significant absorption lines at the quasar emission redshift
(e.g.\ C\,IV, O\,VI).
In previous studies, Broad Absorption Line (BAL) quasars
have been removed from the analysis primarily to prevent 
confusion with intrinsic \ovi\ and/or \nv\ absorption.
We take a less conservative approach
here.  We visually inspected the $1252$ quasars
with $z_{em} > 2.1$ and $r' < 19.5$
to identify quasars with associated absorption.  
In these cases, we limit the DLA search to 
100\AA\ redward of \ovi\ emission and 100\AA\ blueward of 
\lya\ emission. 
However, if BAL contamination is determined to be too severe
the quasar is rejected from further analysis. 

The majority of previous DLA surveys relied on low resolution
`discovery' spectra to first identify DLA candidates.
Follow-up observations were than made of these candidates
to confirm DLA systems and measure their $\N{HI}$ values.
A tremendous advantage of the SDSS spectra is that they have
sufficient resolution to both readily identify DLA candidates
and measure their $\N{HI}$ values.
DLA candidates were identified using an algorithm tuned to the
characteristics of the damped \lya\ profile, in particular
its wide, saturated core.
Our DLA-searching algorithm first 
determines a characteristic signal-to-noise ratio (SNR$_{qso}$) for each quasar
spectrum.  
Ideally, we calculate this value blueward of \lya\ emission 
specifically by taking the median SNR of 150 pixels
lying 51-200 pixels blueward of \lya\ emission.
If the \lya\ emission peak is at less than 200 pixels from the start
of the spectrum, then we calculate SNR$_{qso}$ from the median SNR of
the 150 pixles starting 50 pixels redward of \lya\ emission.
We then define a quantity $n_1$=SNR$_{qso}$/2.5 
restricted to have a value between 1 and 2.
At each pixel $j$ in the spectrum, we then measure the fraction of pixels
with SNR$_j < n_1$ in a window $6 (1+z_j)$ pixels wide 
where $z_j = \lambda_j / 1215.67{\rm \AA} - 1$.
This window was chosen to match the width of the core of a DLA profile
with SDSS spectral resolution and sampling.
Importantly (for fiber data), the
algorithm is relatively insensitive to the effects of
poor sky-subtraction.  
Furthermore, we stress that continuum fitting is unnecessary;
the algorithm works directly on the fluxed data because it focuses
primarily on the core of the damped \lya\ profile.

This algorithm was developed through tests on both
simulated spectra with resolution and SNR comparable to SDSS data
and also on a sub-set of SDSS spectra with known DLA systems.
Our tests indicate that DLA candidates correspond 
to windows where the fraction of pixels
with SNR~$< n_1$ exceeds 60$\%$.
We recorded all regions satisfying this criterion
and reduced them to individual candidates by grouping within
2000~\kms\ bins.
In a sample of 1000 trails on simulated spectra with random
$\N{HI}$ and redshift, we recover 100$\%$ of all DLA systems
with $\log \N{HI} > 20.4$ and all but $\approx 5\%$ of the DLA systems
with $\N{HI} \approx 2 \sci{20} \cm{-2}$.
The algorithm is conservative in that it triggers
many false positive detections, the majority of which are BAL features
or blended \lya\ clouds.
With custom software, it is easy to visually 
identify and account for these cases.

\begin{table}\footnotesize
\begin{center}
\caption{{\sc SDSS QUASAR SAMPLE\label{tab:qso}}}
\begin{tabular}{lccccl}
\tableline
\tableline
Name&$z_{em}$& $z_{start}$& $z_{end}$ & $f_{BAL}^a$ & 
$z_{candidate}$ \\
\tableline\tskip
J094454.24$-00$4330.3&2.292&2.150&2.259&0&\\
J095253.84$+01$1422.1&3.024&2.154&2.984&0&2.204,2.381\\
J100412.88$+00$1257.5&2.239&2.156&2.207&0&\\
J100553.34$+00$1927.1&2.501&2.155&2.466&0&\\
J101014.25$-00$1015.2&2.190&2.143&2.158&0&\\
J101748.90$-00$3124.5&2.283&2.156&2.250&0&\\
J101859.96$-00$5420.2&2.183&2.147&2.151&0&\\
J102606.67$+01$1459.0&2.266&2.157&2.233&0&\\
J102636.96$+00$1530.2&2.178&... & ... &0&\\
J102650.39$+01$0518.3&2.274&2.177&2.192&1&\\
\tableline
\end{tabular}
\end{center}
\tablenotetext{a}{0=No BAL activity; 1=Modest BAL activity, included in analysis; 2=Strong BAL activity, excluded}
\tablecomments{[The complete version of this table is in the electronic edition.]}
\end{table}

Table~\ref{tab:qso} lists the full sample of SDSS-DR1 quasars. 
The columns give the name, $z_{em}$, $z_{start}$,
$z_{end}$, a flag for BAL characteristics, and redshifts
of DLA candidates including the false positive detections.

\begin{figure*}
\begin{center}
\includegraphics[width=6.8in]{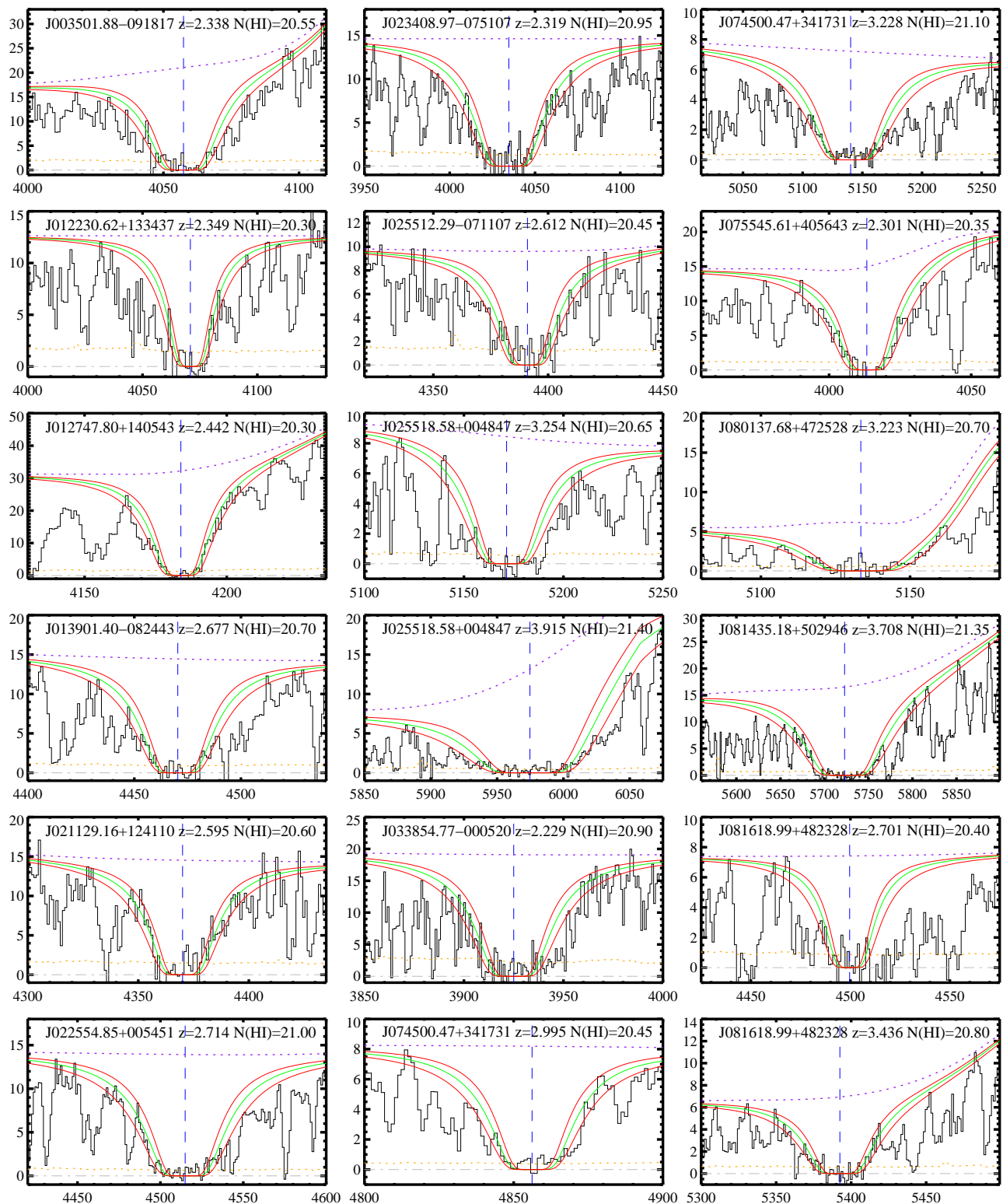}
\figcaption{\lya\ profiles of the \ndla\ damped \lya\ systems comprising
the full statistical sample from the SDSS Data Release 1.
The dotted line traces the assumed continuum of the quasar and the
green solid line is a Voigt profile corresponding to the $\N{HI}$ values
given in Table~\ref{tab:dla}.  All plots have angstroms along the x-axis
and flux $(f_\lambda \times 10^{17}$\,cgs) along the y-axis.
}
\label{fig:data}
\end{center}
\end{figure*}

\begin{figure*}
\begin{center}
\includegraphics[width=6.8in]{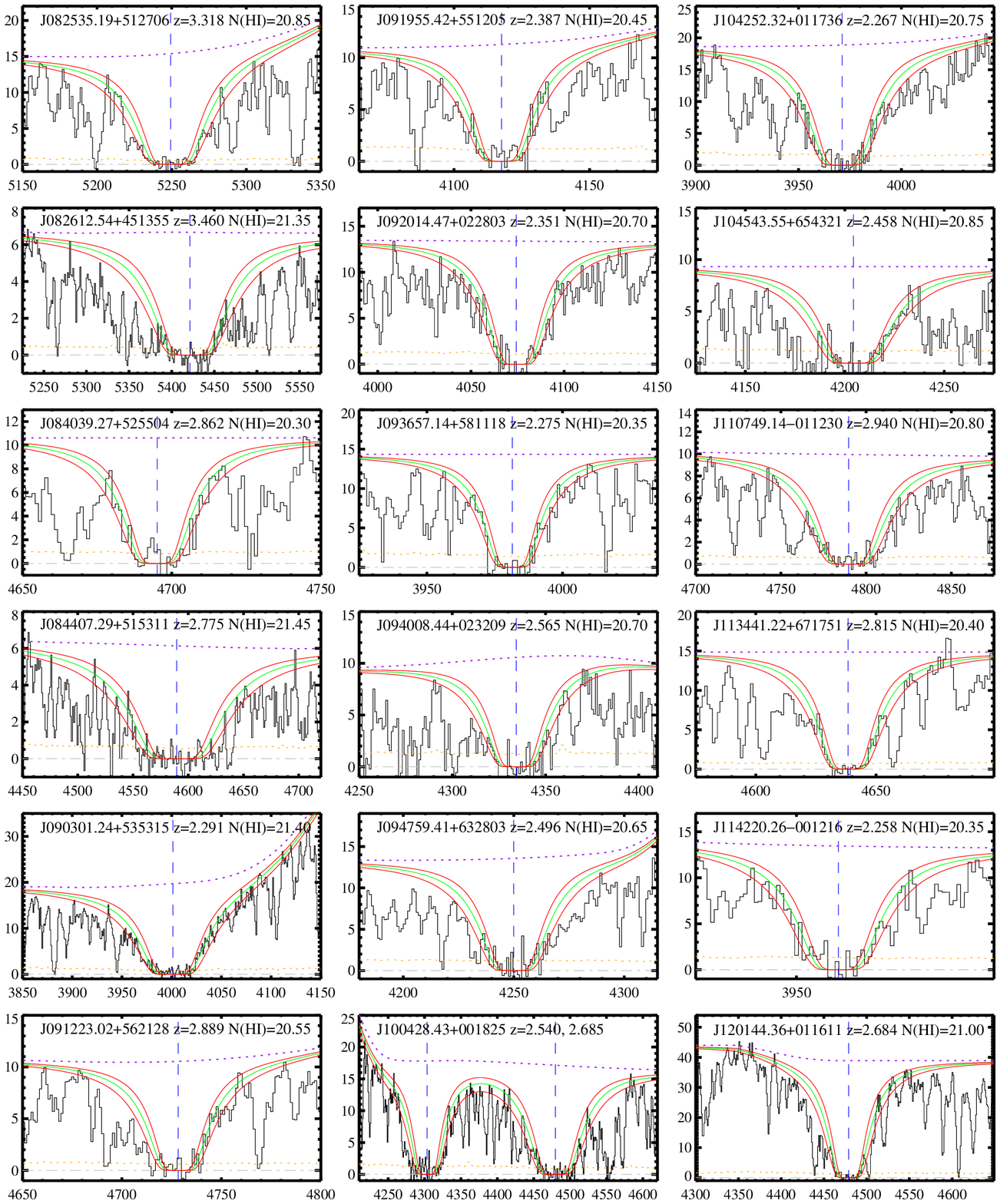}
\end{center}
\end{figure*}

\begin{figure*}
\begin{center}
\includegraphics[width=6.8in]{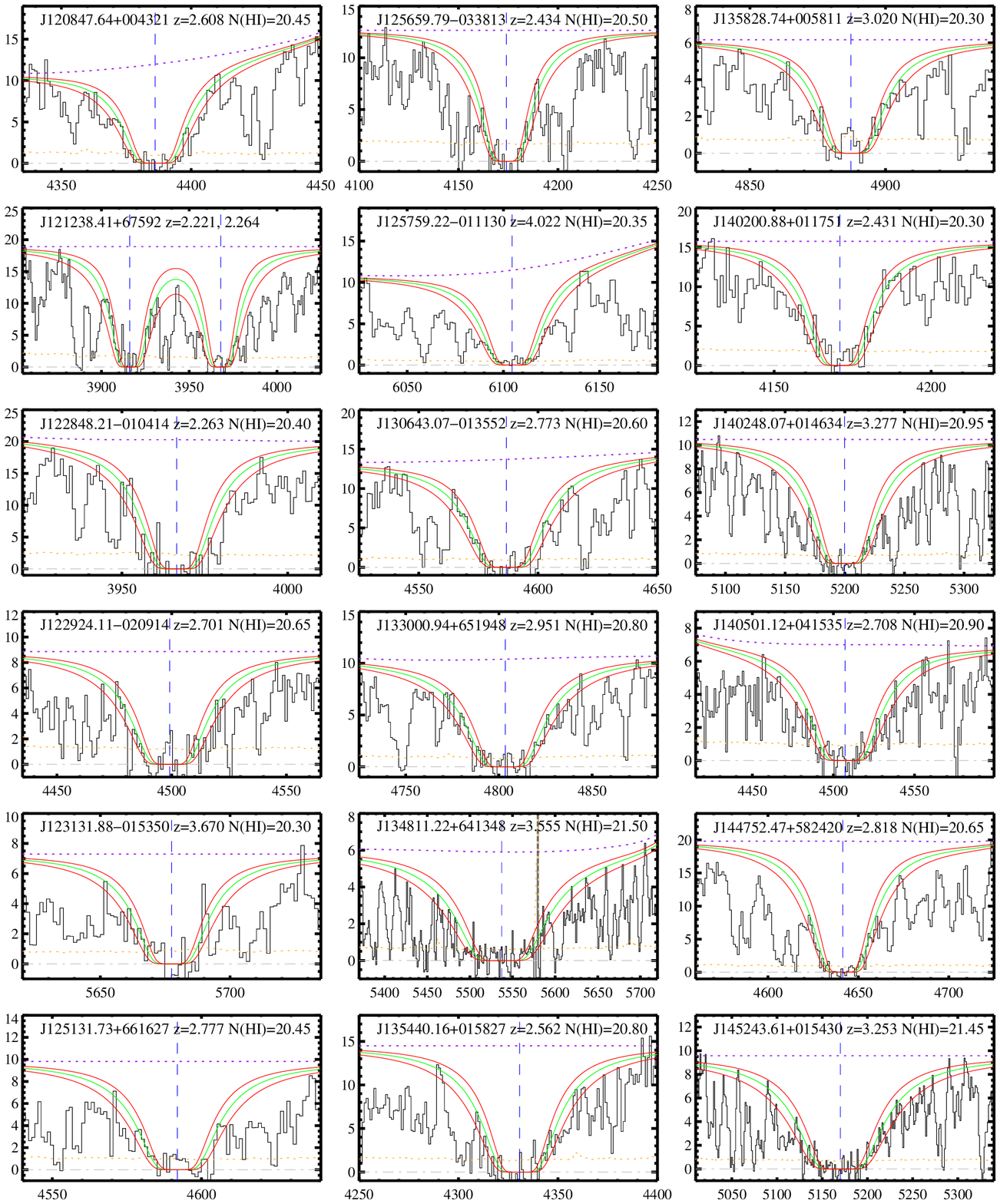}
\end{center}
\end{figure*}

\begin{figure*}
\begin{center}
\includegraphics[width=6.8in]{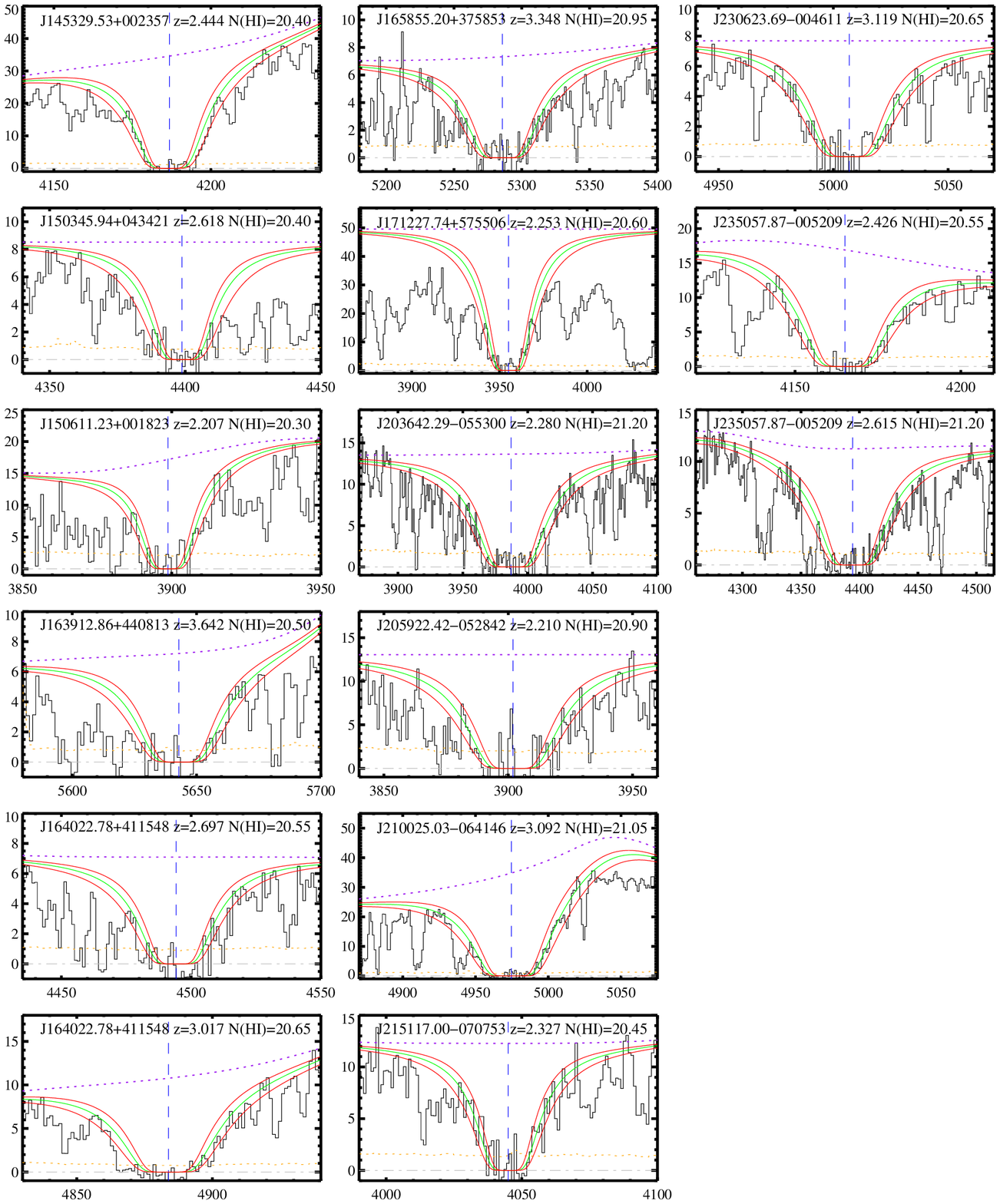}
\end{center}
\end{figure*}

\section{$\N{HI}$ Analysis}

The automated algorithm described in the previous section
triggered 286 DLA candidates. 
We visually inspected the full set of candidates and identified
$\approx 100$ as obvious false positive detections.
For the remainder of the systems, we fit a local continuum 
and a Voigt profile with $FWHM = 2$\,pixels to the data.
The Voigt profile fits to the DLAs quoted in this 
paper are centered on the redshift determined by associated metal-line 
absorption.  Because the metal lines are narrow, these redshifts are
determined precisely.  
As emphasized by \cite{pro03}, the $\N{HI}$ analysis
is dominated by systematic error associated with continuum fitting and line
blending of coincident \lya\ clouds.  The statistical error based 
on a $\chi^2$ minimization routine would be unrealistically low
and largely unmeaningful.  Therefore, we perform a visual fit 
to the data and report a conservative systematic error which we
believe encompasses an interval in $\N{HI}$ corresponding to
a $95\%$~c.l.
For a majority of the profiles, this corresponds to $\pm 0.15$\,dex, 
independent of $\N{HI}$ value.

The \lya\ fits for all \lya\ profiles satisfying $\N{HI} \geq 2 \sci{20} \cm{-2}$
criterion are plotted in Figure 1.  Overplotted in 
each figure are the best fit and our assessment of the error corresponding
to a 95$\%$~c.l.\ interval.
Table~\ref{tab:dla} summarizes the absorption redshift, lists the $\N{HI}$ value
and estimated uncertainty, and gives a brief comment for each profile
(e.g.\ difficult continuum, severe line-blending, poor SNR).

For $\approx 10$ of the DLA systems in the SDSS-DR1 sample, we have
acquired higher resolution spectroscopy ($FWHM \approx 30 \mkms$)
of the \lya\ profile with the Echellette Spectrometer and
Imager \citep{sheinis02} on the Keck~II telescope.
The ESI spectra suffer less from line blending and also
allow for a more accurate determination of the quasar continuum.
Furthermore, several of these systems were observed in previous studies.
We find that our $\N{HI}$ values agree with all previous measurements
to within $0.15$\,dex with no systematic offset.  
Therefore, we are confident in the $\N{HI}$ values reported here.

\begin{table*}\footnotesize
\begin{center}
\caption{{\sc SDSS DLA SAMPLE\label{tab:dla}}}
\begin{tabular}{lccccl}
\tableline
\tableline
Name & $r'$ &$z_{em}$ & $z_{abs}$ & $\log \N{HI}$ & Comment \\
\tableline\tskip
J003501.88$-09$1817&19.10&2.420&2.338&$20.55^{+0.15}_{-0.15}$&continuum\\
J012230.62$+13$3437&19.32&3.010&2.349&$20.30^{+0.15}_{-0.15}$&\\
J012747.80$+14$0543&18.73&2.490&2.442&$20.30^{+0.15}_{-0.15}$&continuum, blending\\
J013901.40$-08$2443&18.68&3.020&2.677&$20.70^{+0.15}_{-0.15}$&\\
J021129.16$+12$4110&18.87&2.950&2.595&$20.60^{+0.15}_{-0.15}$&\\
J022554.85$+00$5451&18.97&2.970&2.714&$21.00^{+0.15}_{-0.15}$&blending\\
J023408.97$-07$5107&18.97&2.540&2.319&$20.95^{+0.15}_{-0.15}$&\\
J025512.29$-07$1107&19.43&2.820&2.612&$20.45^{+0.15}_{-0.20}$&\\
J025518.58$+00$4847&19.27&3.990&3.254&$20.65^{+0.15}_{-0.15}$&continuum, blending\\
&&&3.915&$21.40^{+0.15}_{-0.15}$&continuum, blending\\
J033854.77$-00$0520&18.78&3.050&2.229&$20.90^{+0.15}_{-0.15}$&continuum, blending, poor SNR\\
J074500.47$+34$1731&19.25&3.710&2.995&$20.45^{+0.15}_{-0.15}$&\\
&&&3.228&$21.10^{+0.15}_{-0.15}$&\\
J075545.61$+40$5643&19.23&2.350&2.301&$20.35^{+0.20}_{-0.15}$&blending\\
J080137.68$+47$2528&19.42&3.280&3.223&$20.70^{+0.15}_{-0.15}$&continuum, blending\\
J081435.18$+50$2946&18.34&3.880&3.708&$21.35^{+0.15}_{-0.15}$&\\
J081618.99$+48$2328&19.17&3.570&2.701&$20.40^{+0.20}_{-0.15}$&continuum\\
&&&3.436&$20.80^{+0.15}_{-0.15}$&continuum\\
J082535.19$+51$2706&18.36&3.510&3.318&$20.85^{+0.15}_{-0.15}$&\\
J082612.54$+45$1355&19.23&3.820&3.460&$21.35^{+0.15}_{-0.15}$&blending, poor SNR\\
J084039.27$+52$5504&19.34&3.090&2.862&$20.30^{+0.15}_{-0.15}$&continuum\\
J084407.29$+51$5311&19.44&3.210&2.775&$21.45^{+0.15}_{-0.15}$&continuum\\
J090301.24$+53$5315&18.56&2.440&2.291&$21.40^{+0.15}_{-0.15}$&\\
J091223.02$+56$2128&19.09&3.000&2.889&$20.55^{+0.15}_{-0.15}$&continuum\\
J091955.42$+55$1205&19.02&2.510&2.387&$20.45^{+0.15}_{-0.15}$&continuum\\
J092014.47$+02$2803&19.21&2.940&2.351&$20.70^{+0.15}_{-0.15}$&continuum\\
J093657.14$+58$1118&19.03&2.540&2.275&$20.35^{+0.15}_{-0.15}$&blending\\
J094008.44$+02$3209&19.41&3.220&2.565&$20.70^{+0.15}_{-0.15}$&\\
J094759.41$+63$2803&19.17&2.620&2.496&$20.65^{+0.15}_{-0.15}$&\\
J100428.43$+00$1825&18.50&3.050&2.540&$21.00^{+0.15}_{-0.15}$&\\
&&&2.685&$21.35^{+0.15}_{-0.15}$&\\
J104252.32$+01$1736&18.69&2.440&2.267&$20.75^{+0.15}_{-0.15}$&continuum, poor SNR\\
J104543.55$+65$4321&19.10&2.970&2.458&$20.85^{+0.15}_{-0.15}$&continuum\\
J110749.14$-01$1230&19.22&3.400&2.940&$20.80^{+0.15}_{-0.15}$&blending\\
J113441.22$+67$1751&18.59&2.960&2.815&$20.40^{+0.15}_{-0.15}$&blending\\
J114220.26$-00$1216&18.91&2.490&2.258&$20.35^{+0.15}_{-0.15}$&\\
\tableline
\end{tabular}
\end{center}
\end{table*}

\begin{table*}\footnotesize
\begin{center}
\begin{tabular}{lcccccccc}
& & & & Table 2 -- cont \\
\tableline
Name & $r'$ &$z_{em}$ & $z_{abs}$ & $\log \N{HI}$ & Comment \\
\tableline\tskip
J120144.36$+01$1611&17.53&3.230&2.684&$21.00^{+0.15}_{-0.15}$&blending\\
J120847.64$+00$4321&19.19&2.720&2.608&$20.45^{+0.15}_{-0.15}$&\\
J121238.41$+67$5920&18.68&2.570&2.221&$20.40^{+0.15}_{-0.15}$&\\
&&&2.264&$20.35^{+0.20}_{-0.20}$&\\
J122848.21$-01$0414&18.23&2.660&2.263&$20.40^{+0.15}_{-0.15}$&\\
J122924.11$-02$0914&19.27&3.620&2.701&$20.65^{+0.15}_{-0.15}$&blending\\
J123131.88$-01$5350&19.30&3.900&3.670&$20.30^{+0.15}_{-0.15}$&\\
J125131.73$+66$1627&19.34&3.020&2.777&$20.45^{+0.15}_{-0.15}$&blending\\
J125659.79$-03$3813&19.08&2.970&2.434&$20.50^{+0.15}_{-0.15}$&\\
J125759.22$-01$1130&18.87&4.110&4.022&$20.35^{+0.15}_{-0.15}$&\\
J130643.07$-01$3552&18.82&2.940&2.773&$20.60^{+0.15}_{-0.15}$&\\
J133000.94$+65$1948&18.89&3.270&2.951&$20.80^{+0.15}_{-0.15}$&blending, poor SNR\\
J134811.22$+64$1348&19.12&3.840&3.555&$21.50^{+0.15}_{-0.15}$&\\
J135440.16$+01$5827&19.07&3.290&2.562&$20.80^{+0.15}_{-0.15}$&\\
J135828.74$+00$5811&19.40&3.910&3.020&$20.30^{+0.15}_{-0.15}$&blending\\
J140200.88$+01$1751&18.81&2.950&2.431&$20.30^{+0.15}_{-0.15}$&\\
J140248.07$+01$4634&18.84&4.160&3.277&$20.95^{+0.15}_{-0.15}$&\\
J140501.12$+04$1535&19.31&3.220&2.708&$20.90^{+0.15}_{-0.15}$&poor SNR\\
J144752.47$+58$2420&18.37&2.980&2.818&$20.65^{+0.15}_{-0.15}$&blending\\
J145243.61$+01$5430&18.87&3.910&3.253&$21.45^{+0.15}_{-0.15}$&\\
J145329.53$+00$2357&18.58&2.540&2.444&$20.40^{+0.15}_{-0.15}$&\\
J150345.94$+04$3421&19.49&3.060&2.618&$20.40^{+0.20}_{-0.15}$&blending\\
J150611.23$+00$1823&18.89&2.830&2.207&$20.30^{+0.15}_{-0.15}$&continuum\\
J163912.86$+44$0813&19.22&3.770&3.642&$20.50^{+0.15}_{-0.15}$&\\
J164022.78$+41$1548&19.41&3.080&2.697&$20.55^{+0.15}_{-0.15}$&\\
&&&3.017&$20.65^{+0.15}_{-0.15}$&\\
J165855.20$+37$5853&19.13&3.640&3.348&$20.95^{+0.15}_{-0.15}$&continuum\\
J171227.74$+57$5506&17.46&3.010&2.253&$20.60^{+0.15}_{-0.15}$&blending\\
J203642.29$-05$5300&18.80&2.580&2.280&$21.20^{+0.15}_{-0.15}$&continuum, blending\\
J205922.42$-05$2842&19.01&2.540&2.210&$20.90^{+0.15}_{-0.15}$&continuum, blending, poor SNR\\
J210025.03$-06$4146&18.12&3.140&3.092&$21.05^{+0.15}_{-0.15}$&blending\\
J215117.00$-07$0753&19.26&2.520&2.327&$20.45^{+0.15}_{-0.15}$&continuum\\
J230623.69$-00$4611&19.23&3.580&3.119&$20.65^{+0.15}_{-0.15}$&continuum\\
J235057.87$-00$5209&18.79&3.020&2.426&$20.55^{+0.15}_{-0.15}$&continuum, blending\\
&&&2.615&$21.20^{+0.15}_{-0.15}$&continuum, blending\\
\tableline
\end{tabular}
\end{center}
\end{table*}

\begin{figure}[ht]
\begin{center}
\includegraphics[height=3.6in,angle=90]{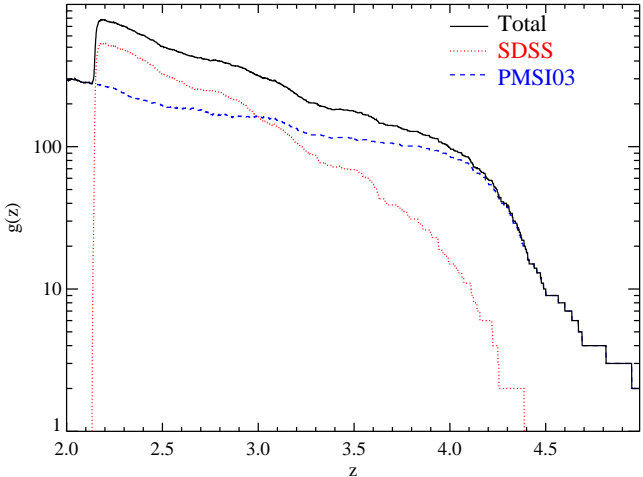}
\figcaption{Redshift path density $g(z)$ as a function of redshift for
the (i) SDSS-DR1 survey (dotted red line); (ii) the PMSI03 compilation
(dashed blue line); and (iii) the combined surveys.
}
\label{fig:gz}
\end{center}
\end{figure}

\section{ANALYSIS}

\subsection{$g(z)$ and $n(z)$}

A simple yet meaningful description of the statistical significance
of any quasar absorption line survey is given by the redshift
path density $g(z)$ \citep[e.g.][]{lzwt91}. 
This quantity corresponds to the number of quasars searched at a given
redshift for the presence of a particular absorption feature, e.g.,
a damped \lya\ system.  We have constructed $g(z)$ for the
SDSS-DR1 sample by implementing the starting and ending redshifts
listed in Table~\ref{tab:qso}.  Figure 2 presents
$g(z)$ for 
(i) the SDSS-DR1 sample (red dotted lines);
(ii) the PMSI03 compilation (dashed blue lines); and
(iii) the combined surveys taking into account overlap between
the two samples (black solid line).
It is evident from Figure 2 that the SDSS-DR1 sample
has greatest statistical impact at $z = 2-3.2$. 
With only $\approx 7\%$ of the projected SDSS database, the SDSS-DR1 
exceeds the redshift path density of the previous two decades of 
research at $z = 2.5$.  
Although the SDSS-DR1 systems have only a modest contribution at
$z>3$, the projected $10\times$ increase
in $g(z)$ for the full SDSS sample promises a major impact for DLA studies
to at least $z=4$. 

\begin{figure}[ht]
\begin{center}
\includegraphics[height=3.6in,angle=90]{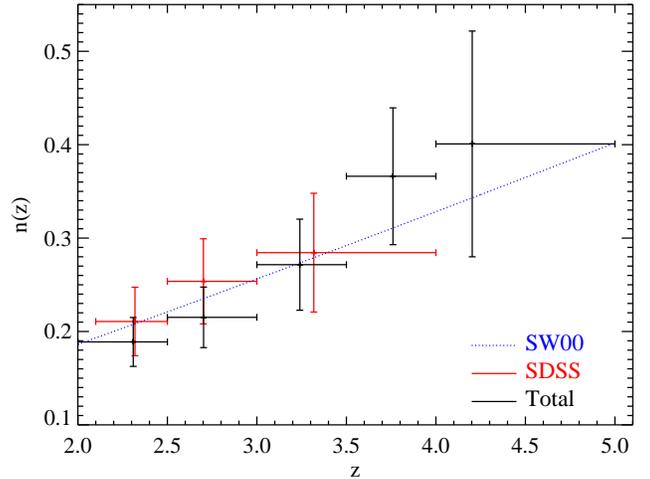}
\figcaption{Incidence of damped \lya\ systems per unit redshift $n(z)$
as a function of redshift for the SDSS-DR1 (red points) and
total samples (black points).  The vertical error bars reflect $1\sigma$
uncertainty assuming Possonian statistics and the horizontal bars
indicate the redshift interval.
The dotted blue line is the fit to $n(z)$ from \cite{storrie00}:
$n(z) = 0.055 (1+z)^{1.11}$.
}
\label{fig:nz}
\end{center}
\end{figure}

Granted a determination of $g(z)$, it is trivial to calculate
the number density of DLA systems per unit redshift
$n(z)$.  Integrating $n(z)$ over several
redshift bins, we derive the results presented in Figure 3 
for the SDSS-DR1 sample (red) and the combined surveys (black).
Overplotted on the figure is the power-law fit to $n(z)$ from
\cite{storrie00}: $n(z) = 0.055 (1+z)^{1.11}$.
The SDSS-DR1 sample is in good agreement with previous analysis; this
bolsters the assertion that our analysis has $>95\%$ completeness.
The combined data sample has uncertainties in $n(z)$ of $10 - 15\%$
for $\Delta z = 0.5$ intervals.  With future SDSS data releases,
we will measure $n(z)$ in $\Delta z = 0.25$ intervals to better
than $5\%$ uncertainty.  This measurement provides an important constraint
on the H\,I cross-section of high redshift galaxies \citep[e.g.][]{nagamine04a}
and thereby models of galaxy formation with CDM cosmology 
\citep[e.g.][]{kauff96,ma97}.
Table~\ref{tab:results} lists the $n(z)$ values for the total sample for the
redshift bins shown in Figure 3.

\clearpage

\begin{figure}[ht]
\begin{center}
\includegraphics[height=3.6in,angle=90]{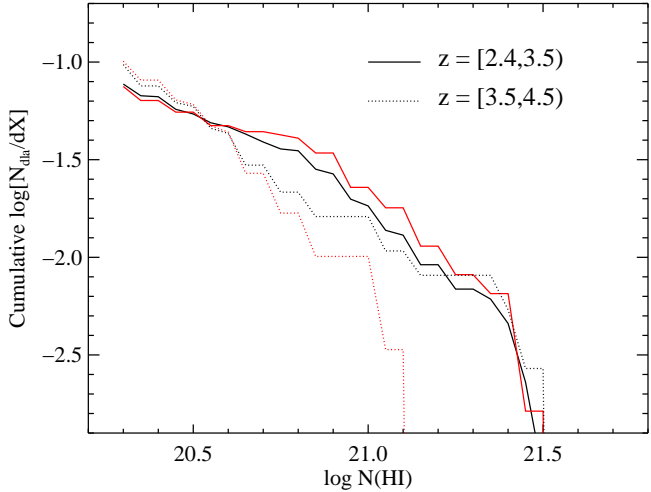}
\figcaption{Cumulative logarithmic incidence of DLA systems per unit
absorption distance interval $dX$ as a function of $\log \N{HI}$.
The red curves correspond to the DLA compilation of PMSI03 and
the black curves refer to the combined sample.
Note that the high redshift results have changed significantly by
including the SDSS-DR1 sample.  
}
\label{fig:cumulf}
\end{center}
\end{figure}

\subsection{$\Omega_g$}

We now turn our attention to the cosmological baryonic mass density in
neutral gas $\Omega_g$ as determined by DLA surveys.
As first described by \cite{wolfe86b}, one can calculate $\Omega_g$ for a given
redshift interval by summing the $\N{HI}$ values of all DLA systems
within that interval and comparing against the total cosmological
distance $\Delta X$ surveyed

\begin{equation}
\Omega_g = \frac{\mu m_H H_0}{c \rho_c} \frac{\Sigma \N{HI}}{\Delta X} \;\; ,
\label{eqn:omegag}
\end{equation}

\noindent where $\mu$ is the mean molecular mass of the gas (taken to be 1.3),
$H_0$ is Hubble's constant, and $\rho_c$ is the critical mass density.
We have calculated $\Delta X$ and $\Omega_g$ for the SDSS-DR1 sample and
the PMSI03 compilation for a $\Omega_m = 0.3$, $\Omega_\Lambda = 0.7$,
$H_0 = 70 \mkms$Mpc$^{-1}$ cosmology consistent with the current
`concordance' cosmology \citep[e.g.][]{spergel03}.

Implicit to Equation~\ref{eqn:omegag} is the presumption that the DLA systems
dominate $\Omega_g$ at all redshift.  A principal result of PMSI03 was
that at $z>3.5$ there are fewer DLA systems with $\N{HI} > 10^{21} \cm{-2}$ and,
therefore, that absorption systems with $\N{HI} \lesssim 10^{20} \cm{-2}$
(the so-called sub-DLA) 
will contribute $\approx 50\%$ of $\Omega_g$.  This point is partially
described by Figure 4 which presents the cumulative
cosmological number density of DLA systems as a 
function of H\,I column density.
The red curves correspond to the compilation analyzed by PMSI03; as
emphasized by these authors there
is a significant drop in the fraction of DLA systems with
large $\N{HI}$ at $z>3.5$ in their compilation.  
The authors then argued 
that the sub-DLA make an important contribution
to $\Omega_g$ at high redshift.
The black lines in Figure 4 correspond to the combined sample.
There is only a modest difference between the PMSI03 and combined samples 
for the $z=[2.4,3.5)$ interval, but at $z>3.5$ (dotted lines)
the SDSS-DR1 results have greatly changed the picture\footnote{We also
note that more accurate $\N{HI}$ measurements from \cite{pro03} indicate that
PMSI03 systematically underestimated several DLA systems with large
$\N{HI}$ value.  These new results are not included in Figure 4,
but are included in the results presented below.}.
Although the SDSS-DR1 systems contribute only 6 new DLA systems at $z>3.5$,
half of these have $\N{HI} > 10^{21} \cm{-2}$.  The resulting cumulative
number density at $z>3.5$ is now in rough agreement with the lower redshift
interval (and the predictions of Nagamine et al.\ 2004a).  
Of course, we suspect the SDSS-DR1 sample shows an abnormally high
fraction of DLA systems at $z>3.5$ with $\N{HI} > 10^{21} \cm{-2}$.
Similarly, we suspect the PMSI03 compilation had disproportionality
too few systems with large $\N{HI}$.  This 
speculation can only be tested through a significantly
larger sample.

We can perform an anlysis similar to PMSI03 to estimate the contribution
of LLS with $\N{HI} = 10^{17.2}$ to $10^{20.3} \cm{-2}$ to $\Omega_g$
in the combined sample.  Adopting their power law fit to the incidence
of LLS $n(z)_{LLS} = 0.07 (1+z)^{2.45}$, one predicts 246.1 LLS with $z>3.5$
for the combined sample where 36 DLA systems are observed.
Assuming the LLS column density distribution follows a power-law
$f(N)_{LLS} = f_0 \N{HI}^\gamma$, we derive $\gamma = -1.31$ and
$f_0 = 10^{4.66}$.  We estimate the contribution of LLS to $\Omega_g$ 
by integrating 

\begin{equation}
\Omega_{LLS} = \frac{\mu m_H H_0}{c \rho_c} 
\int\limits_{10^{17.2}}^{10^{20.3}} N f(N)_{LLS} \, dN = 0.00015 \;\;\; .
\end{equation}
\noindent This value corresponds to $<15\%$ of $\Omega_g$
derived from $z>3.5$ DLA systems (see below).  The fractional contribution is 
3 times lower (and $>4\sigma$ lower) than the results from PMSI03.
It is important to note that this result has large statistical 
and systematic uncertainty.  This includes the parameterization 
of $n(z)_{LLS}$, the
assumed functional form of $f(N)_{LLS}$, and the statistical 
uncertainties in all quantities including $\Omega_g$.  Nevertheless,
we conclude that
there is no longer compelling evidence that Lyman limit systems with
$\N{HI} < 2 \sci{20} \cm{-2}$
contribute significantly to $\Omega_g$ at any redshift.
Given the current uncertainties, however, the exact contribution of the LLS
and DLA systems to $\Omega_g$ will await future studies.

\begin{figure}[ht]
\begin{center}
\includegraphics[height=3.6in,angle=90]{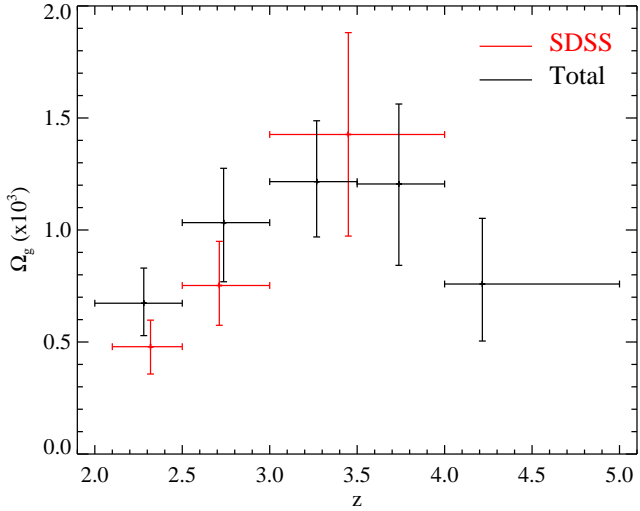}
\caption{Cosmological baryonic mass density in neutral gas $\Omega_g$
as derived from the damped \lya\ systems for the SDSS-DR1 sample
(red points) and the combined sample (black points).
The 1$\sigma$ vertical error bars were derived from a modified bootstrap
analysis described in the text.  Contrary to previous studies, $\Omega_g$
is rising or unchanged to $z=4$ and there is only a statistically
insignificant decline at $z>4$.
}
\label{fig:omega}
\end{center}
\end{figure}

Restricting our analysis of $\Omega_g$ to the DLA systems,
we derive $\Omega_g$ for the SDSS-DR1 sample and the
combined datasets (Figure~\ref{fig:omega}, Table~\ref{tab:results}).
The points plotted in Figure~\ref{fig:omega} are centered
at the $\N{HI}$-weighted redshift in each interval
and the horizontal errors correspond to the redshift bins analyzed.
It is difficult to estimate the error in $\Omega_g$ because
the uncertainty is dominated by sample size, especially the column density
frequency distribution at $\N{HI} > 10^{21} \cm{-2}$.  In the
current analysis, we estimate 1$\sigma$ uncertainties through a modified
bootstrap error analysis.  Specifically, we examine the distribution of $\Omega_g$
values for 1000 trials where we randomly select $m \pm p$ DLA systems for
each redshift interval containing $m$ DLA systems and where 
$p$ is a normally distributed random integer with standard deviation 
$\sqrt{m}$.
The bootstrap technique provides a meaningful assessment of the uncertainty
related to sample size provided the observed dataset samples a significant
fraction of the intrinsic distribution.  At present, we are not confident
that this is the case at any redshift interval, but particularly at 
$z>3$.  The results for the $z>4$ redshift interval are an extreme example
of this concern.  The addition of one or two new 
DLA with $\N{HI} > 10^{21} \cm{-2}$
would significantly increase $\Omega_g$ and its 1$\sigma$ uncertainty.
Therefore, we caution the reader that the 1$\sigma$ errors reported in
Table~\ref{tab:results} likely underestimate the true uncertainty.

\begin{table}\footnotesize
\begin{center}
\caption{{\sc RESULTS\label{tab:results}}}
\begin{tabular}{lccccl}
\tableline
\tableline
Sample &$z$ & N & $n(z)$ & $\Delta X^a$ & 
$\Omega_g (10^{-3})$ \\
\tableline
SDSS \\
& 2.1--2.5&  26&$0.211\pm 0.037$&   505.0&$0.47^{+0.12}_{-0.12}$\\
& 2.5--3.0&  26&$0.254\pm 0.046$&   420.9&$0.76^{+0.20}_{-0.18}$\\
& 3.0--4.1&  19&$0.296\pm 0.065$&   266.3&$1.43^{+0.44}_{-0.45}$\\
Total \\
& 2.0--2.5&  52&$0.189\pm 0.026$&   880.8&$0.67^{+0.16}_{-0.14}$\\
& 2.5--3.0&  44&$0.215\pm 0.032$&   704.5&$1.03^{+0.24}_{-0.26}$\\
& 3.0--3.5&  31&$0.271\pm 0.049$&   421.8&$1.22^{+0.27}_{-0.25}$\\
& 3.5--4.0&  25&$0.366\pm 0.073$&   268.0&$1.21^{+0.36}_{-0.36}$\\
& 4.0--5.0&  11&$0.401\pm 0.121$&   113.5&$0.76^{+0.29}_{-0.26}$\\
\tableline
\end{tabular}
\end{center}
\tablenotetext{a}{Assumes a $\Omega_m = 0.3, \Omega_\Lambda=0.7,H_0=70 \, {\rm km \, s^{-1} \, Mpc^{-1}}$ cosmology.}
\end{table}

 The SDSS-DR1 sample shows no evidence for a
decline in $\Omega_g$ at high redshift;  the results are even suggestive
of an increasing baryonic mass density at $z>3$.
We caution, however, that the uncertainties are large.
Combining the SDSS-DR1 sample with the previous 
studies\footnote{We have updated the measurements presented
in PMSI03 to match the ones presented in \cite{pro03b}.}, 
we reach a similar conclusion except at $z>4$ where 
the current results indicate a drop in $\Omega_g$.
As noted above, the results in the highest redshift interval are very uncertain
owing to small sample size.  
At present, we consider it an open question as to whether $\Omega_g$ declines
at high redshift.  

\begin{figure}[ht]
\begin{center}
\includegraphics[height=3.6in,angle=90]{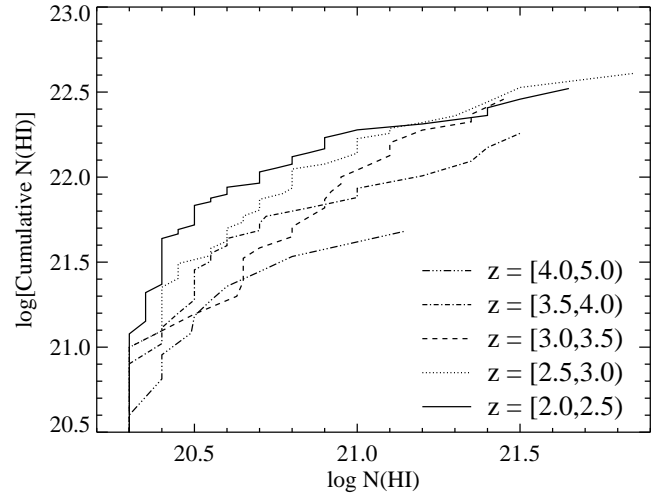}
\caption{Cumulative total $\N{HI}$ as a function of $\log \N{HI}$
for the redshift intervals displayed in Figure~\ref{fig:omega}.  
These curves provide a qualitative assessment of the robustness of the
$\Omega_g$ values to the addition of new DLA systems, especially `outliers'
with large $\N{HI}$.
}
\label{fig:cumul}
\end{center}
\end{figure}

One means of assessing the robustness of the $\Omega_g$ values
to sample size is 
to examine cumulatively the total $\N{HI}$ in the various redshift 
intervals.  This quantity is presented in Figure~\ref{fig:cumul} as a
function of $\N{HI}$ for the combined DLA sample.
On the positive side, the total $\N{HI}$ for the $z<4$ samples
all approach $10^{22.5} \cm{-2}$ which is $\approx 10\times$ larger
than the highest $\N{HI}$ values observed to date.  Therefore, the
results in these intervals are reasonably robust to the inclusion of 
an `outlier' with $\N{HI} \approx 10^{22} \cm{-2}$.
On the other hand, the curves
in Figure~\ref{fig:cumul} demonstrate that 
DLA systems with $\N{HI} > 10^{21} \cm{-2}$ do contribute $\approx 50\%$ 
of the total $\N{HI}$ in each interval.  This point stresses the sensitivity
of $\Omega_g$ to sample size; there are relatively few DLA systems with
$\N{HI} > 10^{21} \cm{-2}$ in each interval. 
Sample variance will be important in any given interval for $\Omega_g$ until 
it includes many systems with $\N{HI} > 10^{21} \cm{-2}$.

\section{SUMMARY AND CONCLUDING REMARKS}

In this paper, we have introduced an automated approach for
identifying DLA systems in the SDSS quasar database.  We
have applied our method to the Data Relase 1 quasar sample and
have identified a statistical sample of \ndla\ DLA systems
including $>50$ previously unpublished cases.
Remarkably, the SDSS 
Data Release 1 exceeds the statistical significance of
the previous two decades of DLA research
at $z \approx 2.5$.  More importantly, this sample
was drawn from a well defined, homogeneous dataset of quasar
spectroscopy.
We present measurements of
the number per unit redshift $n(z)$ of the DLA population
and the contribution of these systems to the cosmological baryonic
mass density in neutral gas $\Omega_g$.  Although the SDSS-DR1
sample does not offer a definitive assessment of either of these
quantities, future SDSS data releases will provide a major
advancement over all previous work.

Our measurements of $n(z)$ are consistent with previous results suggesting
a high completeness level for our DLA survey of the SDSS-DR1.
We find $\Omega_g$ increases with redshift
to at least $z=3$ and is consistent with increasing to $z=4$ and beyond.
This latter claim, however, is subject to significant uncertainty relating
to sample size.
Perhaps the most important result of our analysis is that the full DLA
sample no longer shows significantly fewer DLA systems with large $\N{HI}$ 
at $z>3.5$.  This contradicts the principal result of PMSI03 from their
analysis of the pre-SDSS DLA compilation.  Apparently, their
maximum likelihood approach failed to adequately assess uncertainty
related to sample size.  With the inclusion of only 6 new DLA,
we no longer find that Lyman limit systems with 
$\N{HI} < 2 \sci{20} \cm{-2}$ are required in an analysis of $\Omega_g$.

Before concluding, we offer several additional criticisms of the
PMSI03 analysis and the role of sub-DLA systems.  
First, these authors assumed a three parameter
$\Gamma$-function for the column density frequency distribution of
absorption systems with $\N{HI} > 10^{17.2} \cm{-2}$, \\
$f(N) = (f_*/N_*) (N/N_*)^{-\beta} {\rm e}^{-N/N_*}$.
Although this function gives a reasonable fit to the column density
frequency distribution of the DLA systems, it is not physically 
motivated\footnote{In fact this curve does not smoothly connect to the
power-law derived for quasar absorption 
lines with $\N{HI} < 10^{17.2} \cm{-2}$.}
and, more importantly, places much greater emphasis on sub-DLA than other
functions (e.g.\ a broken power-law).
Future assessments must include other functional forms
to examine this systematic uncertainty. 
Second, the authors did not fit for the normalization of the distribution
function $f_*$.  The uncertainty in this
parameter could easily contribute an additional $>50\%$ to the error budget.
Third, their treatment did not account for sample variance; 
the uncertainties these authors reported were severe underestimates.
Finally (and perhaps most importantly), a recent analysis of a sub-DLA
sample by \cite{dessauges03} has shown that these absorption systems
have very high ionization fractions (see also Howk \& Wolfe 2004 in 
preparation).  Although these absorption systems
may ultimately make an important contribution to the total H\,I mass
density of the universe, they are intrinsically different from the
DLA systems. Indeed, a more appropriate title for this sub-set of
Lyman limit systems is the `super-LLS'.
This gas -- in its present form --
cannot contribute to star formation and is unlikely to be directly
associated with galactic disks or the inner regions of protogalactic
`clumps'.  Any interpretation of results related to the super-LLS
must carefully consider these points \citep[e.g.][]{maller03,peroux03}.

\acknowledgments

We acknowledge the tremendous effort put forth by the SDSS team to
produce and release the SDSS survey.
We thank Art Wolfe, Gabe Prochter, John O'Meara, J. Chris Howk, and Ben Weiner 
for helpful comments and suggestions.
JXP and SHF are partially supported by NSF grant AST-0307408 and
its REU sub-contract.

\end{document}